\documentclass[fleqn,usenatbib]{mnras}

\usepackage{newtxtext,newtxmath}
\usepackage[T1]{fontenc}
\usepackage{ae,aecompl}

\usepackage{graphicx}	% Including figure files
\usepackage{amsmath}	% Advanced maths commands
\usepackage{amssymb}	% Extra maths symbols
\usepackage{booktabs}
\usepackage{threeparttable}
\usepackage[caption=false]{subfig}
\usepackage{array}

\DeclareMathOperator{\sinc}{sinc}

\title[Characterization of 20 LOTAAS pulsars]{The LOFAR Tied-Array All-Sky Survey (LOTAAS): Characterization of 20 pulsar discoveries and their single-pulse behavior}
\author[D. Michilli et al.]{%
D.~Michilli,$^{1,2,3,4}$\thanks{E-mail: danielemichilli@gmail.com}
C.~Bassa,$^{2}$
S.~Cooper,$^{5}$
J.~W.~T.~Hessels,$^{1,2}$
V.~I.~Kondratiev,$^{2,6}$ \newauthor
S.~Sanidas,$^{5,1}$
B.~W.~Stappers,$^{5}$
C.~M.~Tan,$^{5}$
J.~van~Leeuwen$^{2,1}$
I.~Cognard,$^{7,8}$ \newauthor
J.-M.~Grie{\ss}meier,$^{7,8}$
A.~G.~Lyne,$^{5}$
J.~P.~W.~Verbiest,$^{9,10}$
P.~Weltevrede$^{5}$
\\
$^{1}$Anton Pannekoek Institute for Astronomy, University of Amsterdam, Science Park 904, 1098 XH Amsterdam, The Netherlands\\
$^{2}$ASTRON, Netherlands Institute for Radio Astronomy, Oude Hoogeveensedijk 4, 7991 PD, Dwingeloo, The Netherlands\\
$^{3}$Department of Physics, McGill University, 3600 University Street, Montr\'{e}al, QC H3A 2T8, Canada\\
$^{4}$McGill Space Institute, McGill University, 3550 University Street, Montr\'{e}al, QC H3A 2A7, Canada\\
$^{5}$Jodrell Bank Centre for Astrophysics, School of Physics and Astronomy, The University of Manchester, Manchester M13 9PL, UK\\
$^{6}$Astro Space Centre, Lebedev Physical Institute, Russian Academy of Sciences, Profsoyuznaya Str. 84/32, Moscow 117997, Russia\\
$^{7}$LPC2E - Universit\'{e} d'Orl\'{e}ans / CNRS, 45071 Orl\'{e}ans cedex 2, France\\
$^{8}$Station de Radioastronomie de Nan\c{c}ay, Observatoire de Paris, PSL Research University, CNRS, Univ. Orl\'{e}ans, OSUC, 18330 Nan\c{c}ay, France\\
$^{9}$Fakult\"{a}t f\"{u}r Physik, Universit\"{a}t Bielefeld, Postfach 100131, 33501 Bielefeld, Germany\\
$^{10}$Max-Planck-Institut f\"{u}r Radioastronomie, Auf dem H\"{u}gel 69, 53121 Bonn, Germany\\
}

\date{Accepted XXX. Received YYY; in original form ZZZ}

\pubyear{2019}

\begin{document}
\label{firstpage}
\pagerange{\pageref{firstpage}--\pageref{lastpage}}
\maketitle

% Abstract of the paper
\begin{abstract}
We are using the LOw-Frequency ARray (LOFAR) to perform the LOFAR Tied-Array All-Sky (LOTAAS) survey for pulsars and fast transients.
Here we present the astrometric and rotational parameters of 20 pulsars discovered as part of LOTAAS.
These pulsars have regularly been observed with LOFAR at $149$\,MHz and the Lovell telescope at $1532$\,MHz, supplemented by some observations with the Lovell telescope at $334$\,MHz and the Nan\c{c}ay Radio Telescope at $1484$\,MHz.
Timing models are calculated for the 20 pulsars, some of which are among the slowest-spinning pulsars known.
PSR J1236$-$0159 rotates with a period $P \sim 3.6$\,s, while 5 additional pulsars show $P>2$\,s.
Also, the spin-down rates $\dot{P}$ are, on average, low, with PSR J0815$+$4611 showing $\dot{P} \sim 4\times10^{-18}$.
Some of the pulse profiles, generically single-peaked, present complex shapes evolving with frequency.
Multi-frequency flux measurements show that these pulsars have generically relatively steep spectra but exceptions are present, with values ranging between $\sim -4$ and $-1$.
Among the pulsar sample, a large fraction shows large single-pulse variability, with 4 pulsars being undetectable more than $15\%$ of the time and one tentatively classified as a Rotating Radio Transient.
Two single-peaked pulsars show drifting sub-pulses.
\end{abstract}

% Select between one and six entries from the list of approved keywords.
% Don't make up new ones.

%%%%%%%%%%%%%%%%%%%%%%%%%%%%%%%%%%%%%%%%%%%%%%%%%%

%%%%%%%%%%%%%%%%% BODY OF PAPER %%%%%%%%%%%%%%%%%%

\section{Introduction}

Radio pulsars are rotating neutron stars where a small fraction of the spin-down energy powers beamed radio emission that can cross our line-of-sight on every rotation resulting in an observable pulsed signal \citep{pac67,gol68}.
Since their discovery by \citet{hew68}, pulsars have provided a great wealth of scientific discoveries largely thanks to their use as uniquely precise astronomical clocks \citep{man17}.  This has motivated ongoing pulsar surveys at a wide range of radio frequencies, which to date have found close to 3000 sources \citep[ATNF catalog;][]{man05}\footnote{\url{http://www.atnf.csiro.au/people/pulsar/psrcat}}.  
In order to take advantage of these pulsar discoveries, it is necessary to construct a pulsar timing model by measuring the pulses' times-of-arrival (TOAs).
This model describes the rotational and astrometric properties of the pulsar and the propagation of the pulses through the interstellar medium \citep{edw06}.
The pulsar spin and spin-down rate are indicative of the pulsar evolutionary history.
Assuming a simplified model (i.e. dipole braking with constant magnetic field), pulsar parameters can be estimated from these, such as the characteristic age and magnetic field \citep{gol69}.
For this reason, a scatter-plot of pulsar periods $P$ and period derivatives $\dot{P}$ (the so-called $P-\dot{P}$ diagram) provides valuable information on the properties of the pulsar population as a whole.
TOAs for a given pulsar can be obtained directly from its single pulses.
However, the low signal-to-noise ratio (S/N) and erratic shapes of individual  pulses result in larger uncertainties.
Therefore, hundreds of rotational periods are usually added together to increase the S/N and form a stable average pulse profile.
Pulse profiles are correlated with noiseless templates in order to produce TOAs.  
A detailed description of this `timing' procedure is provided by \citet{edw06}.

All pulsars manifest some level of pulse-to-pulse variation in flux and pulse shape. In some cases the emission switches between bi-stable states, and these sources are classified as mode changing pulsars \citep{wan07}.  In other cases, the single pulses form patterns called drifting sub-pulses \citep{tay75}.
Pulsars that are undetected for one or multiple rotational periods are classified as nullers \citep[][here sufficient sensitivity is needed to distinguish between nulling and weak pulses]{bac70}.
The nulling fraction can vary from a small fraction of the rotations to nearly $100$\% \citep[e.g.][]{wan07}.
In the latter case, sources are often discovered through their single pulses and termed Rotating RAdio Transients \citep[RRATs,][]{mcl06}.
These are pulsars whose emission is detected over single rotations separated by large periods of apparent inactivity ($\gtrsim 1$\,minute and up to hours).

Radio waves propagating through the cold plasma present in the interstellar medium (ISM) undergo various propagation effects \citep[e.g][]{ric90}, which are usually more evident for lower-frequency waves ($\lesssim 300$\,MHz).  
These effects are highly relevant both in searching for new pulsars and in measuring precise TOAs.
Dispersion is the frequency-dependence of the wave group velocity. It is quantified by the dispersion measure (DM) and scales as $\nu^{-2}$, where $\nu$ is the observing frequency.  
Diffractive scintillation is the phase perturbation of the waves induced by smaller-scale inhomogeneities in the ISM.
It creates intensity modulations of the signal both in time and frequency.
Diffractive scintillation is typically averaged out by wide-band observations ($> 10$\,MHz) at low-frequencies.
Refractive scintillation is the angular broadening of the radio signal due to larger-scale inhomogeneities in the ISM.
It typically manifests in the time domain as an exponential scattering tail in the pulse profiles, which scales roughly as $\nu^{-4}$.  Scattering can strongly limit the detectability of pulsars at low frequencies because it can wash out the pulsed signal.

The majority of current pulsar surveys are being carried out at frequencies above 300\,MHz, where the sky background brightness is lower and the aforementioned radio propagation effects are less severe \citep{lor04,sto13}.
However, low-frequency observations ($\lesssim 300$\,MHz) present some practical advantages as well and they can probe the pulsar population in a way that complements the view from higher frequencies.
Firstly, the telescope's field-of-view is typically larger at lower frequencies, allowing a larger survey speed for all-sky surveys \citep{sta11}.
Secondly, most pulsars are brighter at lower frequencies \citep{bil16}.
The flux density S of radio pulsars is usually described by a power-law of the observing frequency $\nu$ whose exponent $\alpha$ is called the spectral index ($S\propto\nu^\alpha$).
If a pulsar has a spectrum steeper than the sky background \citep[$\alpha \sim -2.55$,][]{moz17}, it can potentially be detected more easily at lower frequencies -- as long as scattering is modest.
The average spectral index of pulsars is $\alpha = -1.4$, with a standard deviation  of $1$ \citep{bat13}.

Here we report the timing models and other properties of 20 radio pulsars discovered using the Low Frequency ARray \citep[LOFAR,][]{haa13,sta11}, a sensitive radio interferometer that operates at low radio frequencies.
We are using this telescope in the frequency range $119-151$\,MHz to perform the LOFAR Tied-Array All-Sky Survey \citep[LOTAAS,][]{coe14}\footnote{\url{http://www.astron.nl/lotaas}} for pulsars and fast transients in the Northern sky.
A detailed description of the survey is presented in \citet{san18}.
Among the LOTAAS discoveries presented in this paper,
PSR~J0815$+$4611 has been first presented by \citet{jel15}, who identified a steep spectrum, unresolved and polarized point source in continuum images of the 3C\,196 field observed by the LOFAR Epoch of Reionization project (Candi2, Ger de Bruyn, priv. comm.; \citealp{yat13}).
Pulsations were then discovered using a targeted LOFAR beam-formed observation (DDT2\_004, PI: Hessels) and subsequent search over a range of trial DMs (V. Kondratiev, priv. comm.; \citealp{jel15}).
PSR~J1404$+$1159 was discovered by \citet{cha03} and blindly re-detected by LOTAAS.
It did not have a timing model at the time of the LOTAAS re-discovery and we detected it at a very different DM than the value given by \citet{cha03}, so there was initially some ambiguity about whether it was indeed the same source \citep{san18}.
\citet{bri18} recently presented a timing model for the source compatible with the one we obtain.
PSR J0302$+$2252 was first reported by \citet{tyu16}; PSRs J0122+1416, J1635$+$2332 and J2051$+$1248 by \citet{tyu17}; and PSRs J0139$+$3336, J1404$+$1159 and J1848$+$1516 by \citet{tyu18}.
These sources have been blindly detected by LOTAAS around the same time and we present their timing models for the first time.

Pulsars discovered by LOTAAS are regularly monitored using multiple telescopes;  these subsequent timing observations are described in \textsection\ref{sec:observations}.
The timing models obtained for these pulsars are presented in \textsection\ref{sec:timing} and the characteristics of the pulse profiles are described in \textsection\ref{sec:profiles}.
Flux densities and spectral indices are analyzed in \textsection\ref{sec:fluxes}.
Individual sources presenting interesting variations within single observations are further described in \textsection\ref{sec:variations}.
Finally, conclusions are drawn in \textsection\ref{sec:conclusions}.

\section{Observations}\label{sec:observations}

\begin{table*}
\centering
\caption{Number of single detections of pulsars in each frequency band.
Non-detections are indicated with a dash.
Pulsars not observed at a certain frequency are highlighted with an `X'.
The two last columns indicate the total span ranged by the observations and the pulsar names reported by \citet{san18}, before a timing model was available.
}
\label{tab:observations}
\begin{tabular}{lcccccc}
\toprule
PSR & LOFAR & Lovell & NRT & Lovell & Span & Name in  \\
& 149\,MHz & 334\,MHz & 1484\,MHz & 1532\,MHz & months & \citet{san18} \\
\midrule
J0115$+$6325 & 17& X & -- & 41  & 19 & J0115$+$63\\
J0122$+$1416 & 14& -- & -- & -- & 17 & J0121$+$14\\
J0139$+$3336 & 9& -- & X & 31 & 15 & J0139$+$33\\
J0302$+$2252 & 23& 1 & X & 20 & 24 & J0302$+$22\\
J0349$+$2340 & 20& -- & X & -- & 22 & J0349$+$23\\
J0518$+$5125 & 18& -- & X & -- & 21 & J0518$+$51\\
J0742$+$4334 & 17& -- & X & -- & 24 & J0742$+$43\\
J0815$+$4611 & 31& 1 & X & 20 & 33 & J0815$+$4611\\
J0857$+$3349 & 9& 1 & X & -- & 17 & J0857$+$33\\
J1226$+$0005 & 12& 1 & X & -- & 24 & J1226$+$00\\
J1236$-$0159 & 13& 1 & X & 50 & 25 & J1235$-$02\\
J1343$+$6634 & 24& -- & -- & -- & 23 & J1344$+$66\\
J1404$+$1159 & 14& 1 & 11 & 38 & 19 & J1404$+$11\\
J1635$+$2332 & 15& 1 & X & 33 & 16 & J1635$+$23\\
J1735$+$6320 & 25& 1 & X & 50 & 29 & J1735$+$63\\
J1848$+$1516 & 21& 1 & X & 52 & 31 & J1848$+$15\\
J1849$+$2559 & 21& 1 & X & -- & 24 & J1849$+$25\\
J1933$+$5335 & 11& X & X & X & 15 & J1933$+$53\\
J2051$+$1248 & 24& -- & X & 40 & 25 & J2051$+$12\\
J2329$+$4743 & 15& 1 & X & 31 & 15 & J2329$+$47\\
\bottomrule
\end{tabular}
\end{table*}

As discussed in greater detail by \citet{san18}, the LOTAAS survey is performed using the LOFAR `Superterp', a part of the telescope where 6 stations of antennas are closely spaced. 
After a promising candidate is found, its rough sky position is re-observed using the full LOFAR core (up to 24 stations) for confirmation and refined localization.
Thanks to the longer baselines of the full core, the localization improves to roughly arcminute precision.  
The increased sensitivity of the full core compared to the Superterp also means that significantly shorter integrations, typically 15\,minutes, can be used to achieve a S/N sufficient to obtain an adequately precise TOA.
The resulting discoveries are added to the LOFAR timing campaign, where selected pulsars are observed monthly using the full LOFAR core.
All the pulsars presented here have been observed for a span of at least one year; the total set of observations used in this study is reported in Table~\ref{tab:observations}.
Typically, each pulsar is observed for 10 minutes in the timing campaign.
However, due to their weak or sporadic signals, PSRs~J0139$+$3336, J0518$+$5125, J1848$+$1516 and J1236$-$0159, were observed for $15$--$20$ minutes per epoch.
During the timing campaign, pulsars are coherently de-dispersed at the best DM value resulting from the confirmation observation in order to correct for the intra-channel smearing \citep{han75}.
The LOFAR PULsar Pipeline (\textsc{pulp}), an automatic pipeline described by \citet{sta11} and \citet{kon16}, processes the data from the telescope using \textsc{psrchive} \citep{hot04,str12}\footnote{\url{http://psrchive.sourceforge.net}} and \textsc{dspsr} \citep{str11}\footnote{\url{http://dspsr.sourceforge.net}} to produce an \textit{archive} file.
The archives are data-cubes containing the signal folded at the approximate pulsar spin period determined from the initial confirmation observation as a function of phase, polarization, frequency and time.
Full Stokes information is recorded, the $78$\,MHz of available bandwidth is divided into $400$ channels, the pulse phase is divided into $1024$ bins and the time resolution is usually $5$ seconds.
Only for PSRs~J0139$+$3336 and J1848$+$1516 did we store single-pulse-resolved, total intensity archives in order to study their variability over short timescales.

All the pulsars except for PSR J1933$+$5335 were observed with the 76-m Lovell Telescope at an observing frequency of $1532$\,MHz with a bandwidth of $384$\,MHz \citep{bas16}. 
Each pulsar was first observed between 4-5 times in a span of $10$ days, with single observation lasting between $40$ and $60$ minutes. 
If a pulsar was detected, a regular timing campaign began with an average observing cadence 
of two weeks and observing lengths between $6$ and $60$ minutes depending on the detected S/N. 
The data were processed with the digital filter-bank back-end (dfb), which incoherently de-disperses and produces a data-cube containing $1024$ frequency channels, $10$ second-long sub-integrations and $1024$ phase bins.
For the intermittent source PSR J0139$+$3336, the Apollo back-end was used to store single-pulse resolved data. 
These data has a time resolution of $256$\,$\mu$s and $672$ frequency channels of $0.5$\,MHz.

A sample of 18 pulsars were also observed in one occasion with a bandwidth of $64$\,MHz centered at $334$\,MHz with the Lovell Telescope. 
Each pulsar was observed for a duration of $30$ minutes. 
The data were also processed using the dfb with $512$ channels, $10$ second-long sub-integrations and $512$ phase bins.

Four of the pulsars reported have also been observed with the Nan\c{c}ay radio telescope (NRT) using the NUPPI back-end with a bandwidth of $512$\,MHz centered around $1484$\,MHz.
The data, which are coherently de-dispersed, have a frequency resolution of $4$\,MHz, sub-integrations with a duration of $15$ seconds and $2048$ phase bins.
Typical observation durations were between $\sim 10$ and $40$ minutes.
While PSRs J0115$+$6325 and J0122$+$1416 have been observed two times each, PSR J1343$+$6634 has been observed 9 times without detecting the source.

\section{Timing models}\label{sec:timing}
LOFAR observations supplemented by Lovell TOAs, when available, have been used to construct the timing models presented here.
For each pulsar, initial timing parameters were obtained by using \textsc{presto} \citep{ran01}\footnote{\url{https://www.cv.nrao.edu/~sransom/presto}} to maximize the S/N of the pulse profile in the confirmation observation.
This resulted in approximate values for the period and DM of the sources.
The period derivative was initially set to zero.
The position determined by maximizing the pulse profile S/N of the beam grid of the confirmation observation \citep{san18} was used as a starting point in the timing model.

LOFAR TOAs were obtained by using standard pulsar timing methods.
The \textsc{paz} utility from the \textsc{psrchive} package and \textsc{clean.py} from \textsc{coast guard} \citep{laz16}\footnote{\url{https://github.com/plazar/coast\_guard}} were used to automatically remove radio frequency interference (RFI) present in the observations.
After a visual inspection of the data to remove additional RFI or corrupted observations, sub-integrations and channels were summed to obtain a single pulse profile for each observation.
Since PSRs~J0139$+$3336 and J1848$+$1516 show only very sporadic radio pulses, specific periods where the sources were active have been manually extracted for these pulsars.  
An analytic pulse profile template was generated for each pulsar by fitting the profile having the highest S/N with von Mises functions using \textsc{paas} from \textsc{psrchive}.
For each pulsar, \textsc{pat} has been used to cross-correlate the observed profiles with this analytic template in order to obtain one TOA per observation.
LOFAR TOAs are referenced to the topocentric position of the center of the LBA CS002 station.
For the 11 pulsars detected by the Lovell telescope at $1532$\,MHz, a single pulse profile and a TOA were generated from each observation following \citet{hob04}.

The TOAs obtained for each pulsar were fitted with \textsc{tempo2} \citep{hob06}\footnote{\url{https://bitbucket.org/psrsoft/tempo2}} using the initial timing model.
Most of the pulsars which were detected with the Lovell telescope at $1532$\,MHz were observed at high cadence, allowing us to resolve any phase ambiguities and get an initial coherent timing model.
However, since \textsc{tempo2} can only fit TOAs with known integer rotation counts in between \citep{fre18}, for pulsars only detectable at LOFAR the cadence of our observations meant that in some cases the phase ambiguities could not be resolved. 
We therefore used a brute-force algorithm in these cases.
This algorithm fitted a set of initial spin periods around the value from the discovery observation, producing a series of plots with the relative residuals.
In this way, assuming that the other parameters have a much smaller impact on the fit, it is possible to identify a more precise spin period.
Using this simple yet effective technique, a timing model could be obtained for all the pulsars in the sample.

\begin{table*}
\centering
\caption{Parameters of the timing models for the 20 pulsars presented. The different columns report the source name, Right Ascension and Declination referenced to the J2000 frame, the reference epoch of the parameters, the measured values of spin period, its time derivative and DM.
Values in parentheses are 1-$\sigma$ uncertainties on the last digit. 
}
\label{tab:ephemeris}
\begin{tabular}{lllllll}
\toprule
PSR & RA & DEC & Epoch & $P$ & $\dot{P}$ & DM \\
(J2000) & (J2000) & (J2000) & (MJD) & (s) & (10$^{-15}$) & (pc\,cm$^{-3}$) \\
\midrule
J0115$+$6325 &  01:15:45.87(1) &        $+$63:25:50.8(1) &      57876 & 0.521455427473(5) &     1.599(3) &      65.069(4) \\
J0122$+$1416 &  01:22:01.31(3) &        $+$14:16:17(1) &        57876 & 1.38899395038(2) &      3.803(1) &      17.693(3) \\
J0139$+$3336 &  01:39:57.23(4) &        $+$33:36:59.7(9) &      57901 & 1.2479609557(1) &       2.064(8) &      21.23(1) \\
J0302$+$2252 &  03:02:31.990(4) &       $+$22:52:12.1(2) &      57811 & 1.207164839778(2) &     0.0825(1) &     18.9922(6) \\
J0349$+$2340 &  03:49:57.38(4) &        $+$23:40:53(2) &        57813 & 2.42077097760(4) &      1.0995(7) &     62.962(5) \\
J0518$+$5125 &  05:18:26.145(8) &       $+$51:25:58.7(2) &      57829 & 0.912511685262(8) &     0.191(1) &      39.244(2) \\
J0742$+$4334 &  07:42:42.13(1) &        $+$43:34:02.2(3) &      57998 & 0.606190680037(6) &     0.371(2) &      36.255(4) \\
J0815$+$4611 &  08:15:59.4623(8) &      $+$46:11:53.24(2) &     57662 & 0.4342422517307(2) &    0.0039(1) &     11.2738(3) \\
J0857$+$3349 &  08:57:07.07(2) &        $+$33:49:17.0(6) &      57876 & 0.242961077060(4) &     0.24(1) &       23.998(3) \\
J1226$+$0005 &  12:26:14.4(2) & $+$00:05:44(7) &        57785 & 2.28507617026(8) &      2.474(2) &      18.50(1) \\
J1236$-$0159 &  12:36:02.5(6) & $-$01:59:10(20) &       57803 & 3.5975735967(2) &       5.103(2) &      19.08(3) \\
J1343$+$6634 &  13:43:59.26(2) &        $+$66:34:25.05(9) &     57785 & 1.39410378554(1) &      1.0882(7) &     30.031(3) \\
J1404$+$1159 &  14:04:36.987(8) &       $+$11:59:16.0(2) &      57820 & 2.65043929892(4) &      1.3768(6) &     18.499(4) \\
J1635$+$2332 &  16:35:05.36(1) &        $+$23:32:23.2(3) &      57935 & 1.20869424580(2) &      0.863(5) &      37.568(6) \\
J1735$+$6320 &  17:35:06.562(4) &       $+$63:20:00.06(3) &     57760 & 0.510718135408(1) &     0.3226(4) &     41.853(1) \\
J1848$+$1516 &  18:48:56.13(2) &        $+$15:16:44.1(4) &      57655 & 2.23376977466(5) &      1.6813(8) &     77.436(9) \\
J1849$+$2559 &  18:49:47.555(1) &       $+$25:59:57.66(2) &     57785 & 0.5192634055906(6) &    0.1798(3) &     75.0016(4) \\
J1933$+$5335 &  19:33:01.1(1) & $+$53:35:43(1) &        57546 & 2.052574490(4) &        1.26(2) &       33.54(3) \\
J2051$+$1248 &  20:51:29.66(2) &        $+$12:48:21.5(6) &      57811 & 0.55316745256(2) &      0.019(6) &      43.45(1) \\
J2329$+$4743 &  23:29:31.548(7) &       $+$47:43:39.73(6) &     57950 & 0.728408609085(4) &     0.016(2) &      44.012(2) \\
\bottomrule
\end{tabular}
\end{table*}

In order to obtain more precise values of the pulsar DMs, LOFAR observations were split into two frequency sub-bands and a TOA was calculated for each one.
The same templates were used for the two sub-bands; the good precision of the obtained timing models (see below) justifies this choice.
In the analysis, we did not include possible DM or profile variations over time that are sometimes detected at low frequencies \citep[e.g.][]{don19,mic18}.
The low frequency of LOFAR observations and the availability of the $1532$\,MHz observations with Lovell for most of the pulsars allowed DM uncertainties $\lesssim 0.01$\,pc\,cm$^{-3}$ (see Table~\ref{tab:ephemeris}).
In all cases, the TOAs from the LOFAR and Lovell telescopes were well described by the same timing model after fitting an arbitrary jump in phase between the two instruments to account for a possible phase offset due to different cable length and differences in the reference pulse phase of the template profile.
Both the observatory clocks are referenced to the GPS time system.

The timing models, obtained by using the solar system ephemeris model DE405 \citep{sta98} and performing an unweighted fit, are reported in Table~\ref{tab:ephemeris}.
Some of the pulsars presented here are among those with the slowest periods ever measured \citep{man05}.
It is interesting to note that PSR J0250$+$5854, the slowest radio pulsar ever found, was also discovered by LOTAAS (\citealp{tan18}; see also \citealp{san18} for a discussion of LOTAAS sensitivity to slow pulsars).
The pulsars in the sample presented here also have, on average, low values of $\dot{P}$.

\begin{figure}
\centering
\includegraphics{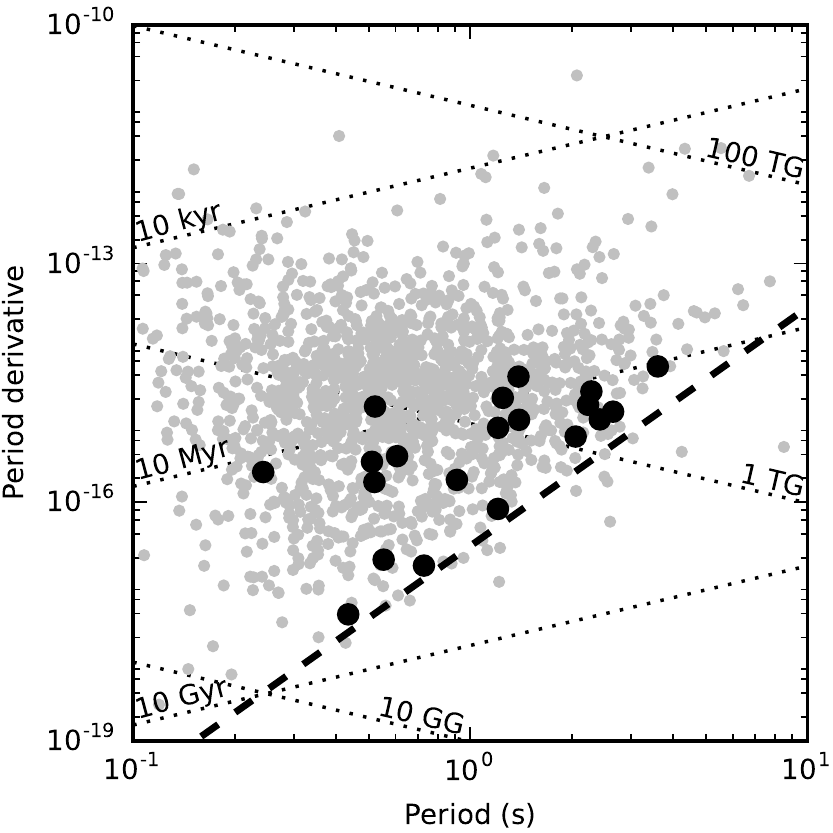}
\caption{
$P-\dot{P}$ diagram of the new pulsar discoveries represented with black dots.
Grey dots represent non-recycled, `slow' radio pulsars \citep[ATNF catalog;][]{man05}.
Dotted lines represent the indicated characteristic ages and magnetic fields, while the dashed line represents the `classical' death line below which dipolar rotators are not expected to emit radio signals \citep{rud75}.
}
\label{fig:ppdot}
\end{figure}

In Fig.~\ref{fig:ppdot}, the new pulsar discoveries are plotted on the $P-\dot{P}$ diagram together with known normal, non-recycled radio pulsars \citep[from the ATNF catalog;][]{man05}.
The relatively high P and low $\dot{P}$ of the sample imply that the new pulsars are on average closer to the death line than the majority of the pulsar population.
It is unclear whether this is due to survey observational selection effects or if it is a real effect, e.g. due to older pulsars having on average steeper radio spectra.
This will be further investigated in a future study using the full sample of LOTAAS discoveries.

\begin{table*}
\centering
\begin{threeparttable}
\caption{Quantities derived from the timing parameters presented in Table~\ref{tab:ephemeris} assuming dipole braking with constant magnetic fields, short initial periods and a moment of inertia $I=10^{45}$\,g\,cm$^2$.
$gl$ and $gb$ are the galactic coordinates, $t_c$ is the characteristic age, $B$ is the surface magnetic field, $\dot{E}$ is the spin-down energy and d the distance of the pulsars.
}
\label{tab:derived}
\begin{tabular}{llllllll}
\toprule
PSR & $gl$ & $gb$ & $\log t_c$ & $\log B$ & $\log\dot{E}$ & d$^\text{a}$ & d$^\text{b}$\\
 & (deg) & (deg) & (yr) & (G) & (erg\,s$^{-1}$) & (kpc) & (kpc)\\
\midrule
J0115$+$6325 &  125.65 &        0.69 &  6.7 &   12.0 &  32.6 &  2.2 &   1.9 \\
J0122$+$1416 &  134.03 &        -47.94 &        6.8 &   12.4 &  31.7 &  0.8 &   1.6 \\
J0139$+$3336 &  134.38 &        -28.17 &        7.0 &   12.2 &  31.6 &  1.0 &   1.5 \\
J0302$+$2252 &  158.44 &        -30.82 &        8.4 &   11.5 &  30.3 &  0.7 &   1.0 \\
J0349$+$2340 &  167.43 &        -23.38 &        7.5 &   12.2 &  30.5 &  3.3 &   3.7 \\
J0518$+$5125 &  158.26 &        7.90 &  7.9 &   11.6 &  31.0 &  1.4 &   1.3 \\
J0742$+$4334 &  175.54 &        27.21 & 7.4 &   11.7 &  31.8 &  1.3 &   1.6 \\
J0815$+$4611 &  173.63 &        33.45 & 9.3 &   10.6 &  30.3 &  0.4 &   0.4 \\
J0857$+$3349 &  190.14 &        39.74 & 7.2 &   11.4 &  32.8 &  0.9 &   1.6 \\
J1226$+$0005 &  289.28 &        62.30 & 7.2 &   12.4 &  30.9 &  0.9 &   1.9 \\
J1236$-$0159 &  295.06 &        60.65 & 7.0 &   12.6 &  30.6 &  0.9 &   2.0 \\
J1343$+$6634 &  114.90 &        49.73 & 7.3 &   12.1 &  31.2 &  1.8 &   < 13.8 \\
J1404$+$1159 &  355.08 &        67.11 & 7.5 &   12.3 &  30.5 &  1.4 &   2.2 \\
J1635$+$2332 &  42.00 & 39.75 & 7.3 &   12.0 &  31.3 &  4.8 &   < 15.7 \\
J1735$+$6320 &  92.72 & 32.55 & 7.4 &   11.6 &  32.0 &  3.2 &   < 19.1 \\
J1848$+$1516 &  46.33 & 7.44 &  7.3 &   12.3 &  30.8 &  3.3 &   3.5 \\
J1849$+$2559 &  56.24 & 11.87 & 7.7 &   11.5 &  31.7 &  3.9 &   6.1 \\
J1933$+$5335 &  85.59 & 15.76 & 7.4 &   12.2 &  30.8 &  2.2 &   2.5 \\
J2051$+$1248 &  59.36 & -19.45 &        8.7 &   11.0 &  30.7 &  2.5 &   4.1 \\
J2329$+$4743 &  108.96 &        -12.91 &        8.9 &   11.0 &  30.2 &  2.2 &   2.4 \\
\bottomrule
\end{tabular}
\begin{tablenotes}
\item[a] Value based on the NE2001 electron density model \citep{cor02}.
\item[b] Value based on the YMW16 electron density model \citep{yao17}.
\end{tablenotes}
\end{threeparttable}
\end{table*}

Physical quantities were derived from the timing model parameters by using standard assumptions.
The characteristic age, dipole magnetic field strength and spin-down energy of the pulsars \citep{lor04} are reported in Table~\ref{tab:derived}.
Also reported in the table are the pulsar distances derived from their DMs using both the NE2001 \citep{cor02} and the YMW16 \citep{yao17} models for the free electron density distribution in the Milky Way.
The latter model implies a maximum expected Galactic contribution lower than the value measured for three pulsars (PSRs~J1343$+$6634, J1635$+$2332 and J1735$+$6320), indicating possible improvements needed in the model (see the discussion in \citealp{san18}, which include a larger pulsar sample).
Given the relatively long rotation period and short observing timespan of the pulsars presented here, it was not possible to obtain reliable proper motion values since they did not affect the residuals significantly.

\begin{figure*}
\centering
\includegraphics{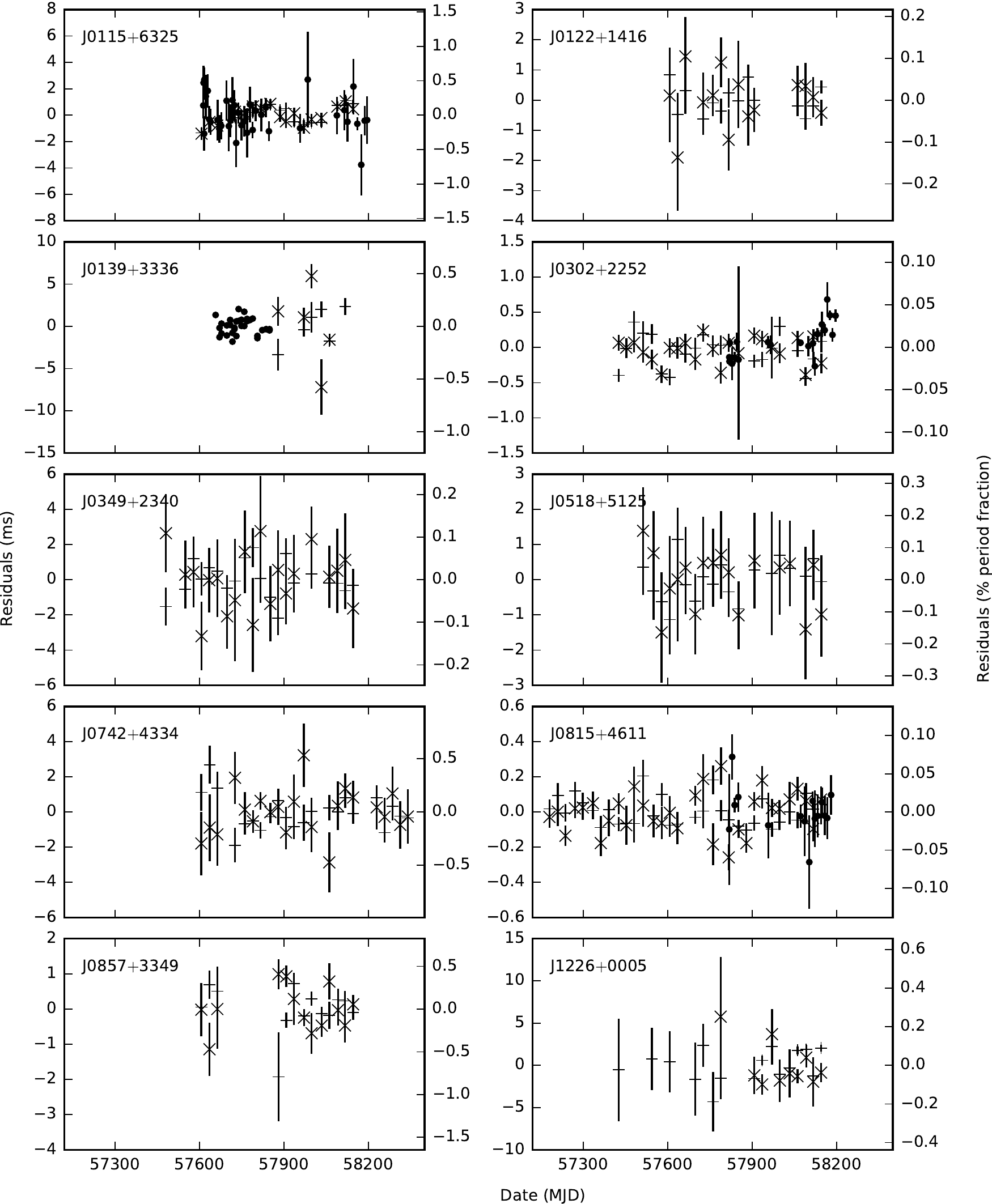}
\caption{
Timing residuals of the models presented in Table~\ref{tab:ephemeris}.
Different symbols represent different observing frequencies, with `+' for LOFAR lower band ($\sim 130$\,MHz), `x' for LOFAR higher band ($\sim 170$\,MHz) and dots for Lovell at $1532$\,MHz.
}
\label{fig:residuals}
\end{figure*}

\begin{figure*}
\ContinuedFloat
\centering
\includegraphics{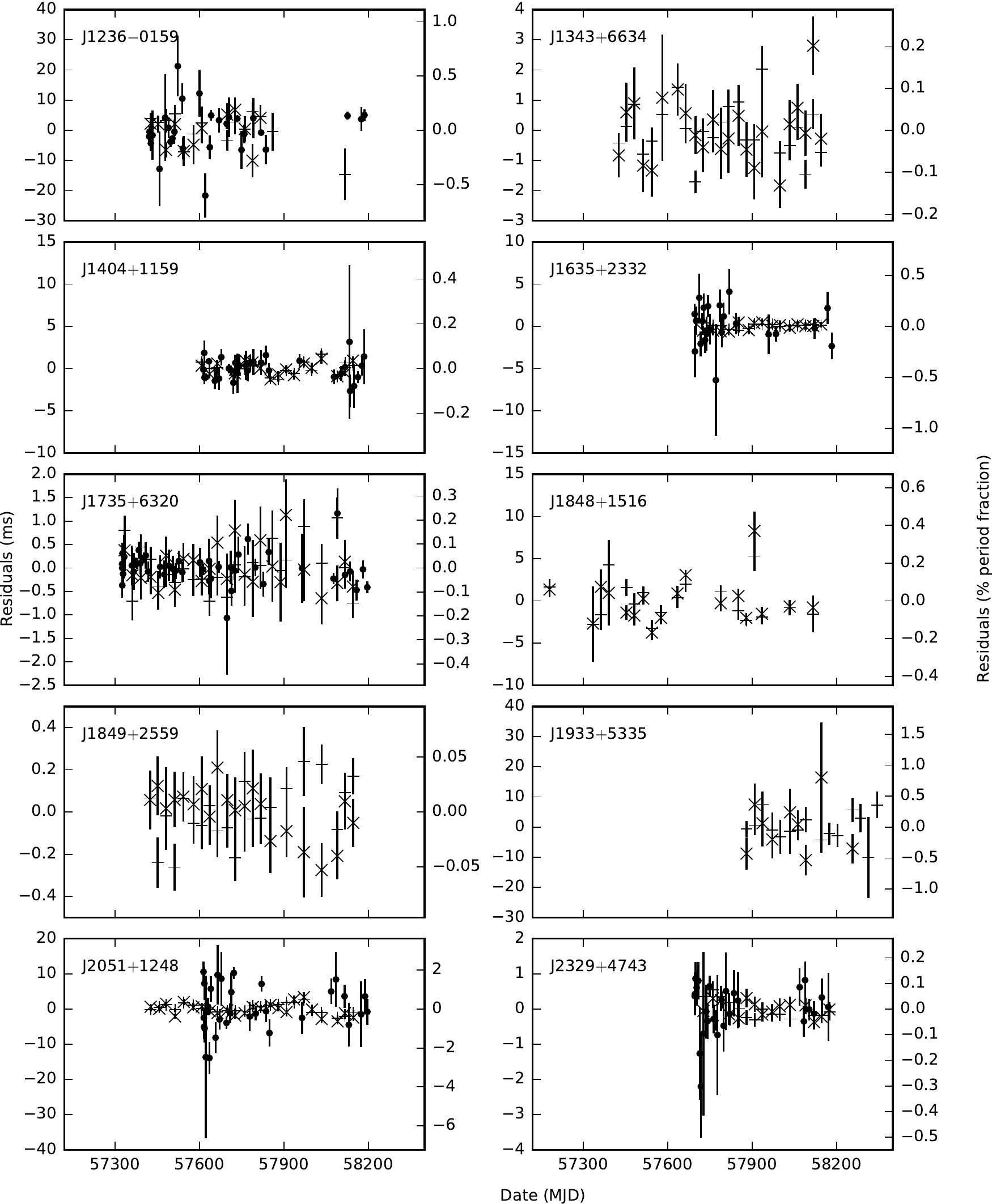}
\caption{
\textit{Continued from previous page.}
}
\end{figure*}

The timing residuals, i.e.\ the difference between observed and model-predicted TOAs, are shown in Fig.~\ref{fig:residuals}.
The long period and sometimes irregular emission of some of the sources (see discussion in \textsection\ref{sec:variations}) imply that the integrated pulse profile for some of the observations might be formed by too few single pulses to stabilize.
This would contribute to the scatter in the TOAs.
However, the timing precision achieved is on average higher than what it is often obtained on similar timespans, and a few pulsars are particularly good timers compared to typical slow pulsars (e.g.\ \citealp{hob04}; the residuals of PSRs~J0302$+$2252, J0815$+$4611 and J1849$+$2559 have a root mean square lower than $200$\,$\mu$s).
This relatively high precision could be due to the narrowness of the peaks in the pulse profiles (\textsection\ref{sec:profiles}) and the low impact of timing noise for these pulsars \citep[e.g.][]{hob04}.

\section{Pulse profiles}\label{sec:profiles}

\begin{figure*}
\centering
\includegraphics{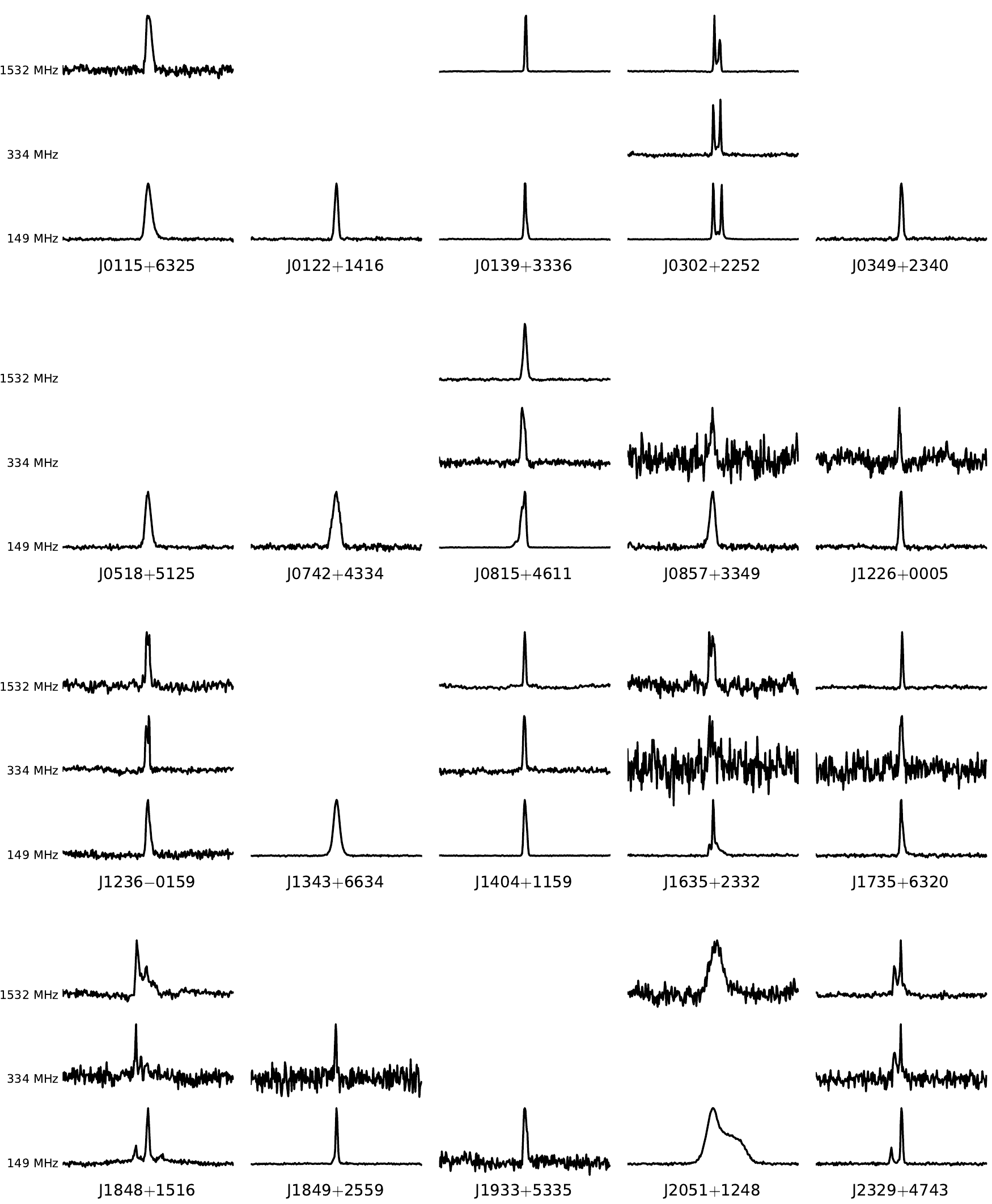}
\caption{Cumulative pulse profiles of the pulsars presented here.
Pulse peaks are all normalized to the same height and full rotational phase windows are shown.
The profiles at different frequencies have been aligned by applying the timing models presented in Table~\ref{tab:ephemeris} and then rotated to show the main peak at the center for the $149$-MHz profile.
}
\label{fig:profiles}
\end{figure*}

We obtained a refined pulse profile from each observation by applying the timing models presented in Table~\ref{tab:ephemeris}.
For each pulsar and observing frequency, the profiles from all observations have been added together to form a global profile; these are presented in Fig.~\ref{fig:profiles}.
The profiles are stored in the EPN Database of Pulsar Profiles\footnote{\url{http://www.epta.eu.org/epndb/}} where they can be accessed on-line.
Since the flux of PSR~J0139$+$3336 is highly variable, rotations where the pulsar was active have been manually selected before forming the integrated profile.

Most pulse profiles are single-peaked, with the exceptions of PSRs~J0302+2252 and J2329+4743, which are double-peaked.
In addition, the single peaks of PSRs~J0815+4611, J1236-0159, J1635+2332, J1848+1516 and J2051+1248 have complex shapes, while the rest are well-fitted by a single von Mises function.
The features of the profiles can be seen in detail in the on-line version stored in the EPN database. 

\begin{table*}
\centering
\caption{Characteristics of the pulse profiles shown in Fig.~\ref{fig:profiles}. W is the width at a fraction of the peak intensity and $\delta$ is the duty cycle. The subscripts indicate the percentage of the peak intensity that the value refers to.
Uncertainties are $\sim 1$\,ms on the width values and $\sim 0.1\%$ on the duty cycle values.
}
\label{tab:profiles}
\begin{tabular}{lcccccccccccccccc}
\toprule
PSR && \multicolumn{3}{c}{$W_{20}$ (ms)} && \multicolumn{3}{c}{$\delta_{20}$ (\%)} && \multicolumn{3}{c}{$W_{50}$ (ms)} && \multicolumn{3}{c}{$\delta_{50}$ (\%)} \\
  & & 149 & 334 & 1532  & & 149 & 334 & 1532  & & 149 & 334 & 1532  & & 149 & 334 & 1532 \\
 & & MHz & MHz & MHz  & & MHz & MHz & MHz  & & MHz & MHz & MHz  & & MHz & MHz & MHz \\
\midrule
J0115$+$6325 & & 35& --& 28 & & 6.7& --& 5.3 & & 22& --& 19 & & 4.3& --& 3.7 \\
J0122$+$1416 & & 48& --& -- & & 3.4& --& -- & & 32& --& -- & & 2.3& --& -- \\
J0139$+$3336 & & 32& --& 22 & & 2.6& --& 1.8 & & 15& --& 14 & & 1.2& --& 1.2 \\
J0302$+$2252 & & 79& 66& 55 & & 6.6& 5.4& 4.6 & & 70& 59& 47 & & 5.8& 4.9& 3.9 \\
J0349$+$2340 & & 71& --& -- & & 2.9& --& -- & & 50& --& -- & & 2.0& --& -- \\
J0518$+$5125 & & 51& --& -- & & 5.5& --& -- & & 33& --& -- & & 3.7& --& -- \\
J0742$+$4334 & & 44& --& -- & & 7.3& --& -- & & 30& --& -- & & 5.0& --& -- \\
J0815$+$4611 & & 21& 17& 16 & & 4.8& 3.8& 3.7 & & 15& 12& 9 & & 3.4& 2.8& 2.2 \\
J0857$+$3349 & & 14& 14& -- & & 5.6& 5.6& -- & & 9& 5& -- & & 3.5& 2.1& -- \\
J1226$+$0005 & & 69& 47& -- & & 3.0& 2.1& -- & & 48& 32& -- & & 2.1& 1.4& -- \\
J1236$-$0159 & & 150& 110& 176 & & 4.2& 3.1& 4.9 & & 94& 96& 90 & & 2.6& 2.7& 2.5 \\
J1343$+$6634 & & 91& --& -- & & 6.5& --& -- & & 57& --& -- & & 4.1& --& -- \\
J1404$+$1159 & & 81& 65& 55 & & 3.0& 2.5& 2.1 & & 58& 45& 36 & & 2.2& 1.7& 1.4 \\
J1635$+$2332 & & 38& 57& 53 & & 3.2& 4.7& 4.4 & & 14& 32& 46 & & 1.1& 2.6& 3.8 \\
J1735$+$6320 & & 15& 16& 9 & & 3.0& 3.0& 1.7 & & 9& 10& 6 & & 1.7& 2.0& 1.1 \\
J1848$+$1516 & & 208& 106& 261 & & 9.3& 4.7& 11.7 & & 40& 20& 146 & & 1.8& 0.9& 6.5 \\
J1849$+$2559 & & 11& 17& -- & & 2.1& 3.2& -- & & 6& 5& -- & & 1.3& 1.0& -- \\
J1933$+$5335 & & 67& --& -- & & 3.3& --& -- & & 53& --& -- & & 2.6& --& -- \\
J2051$+$1248 & & 147& --& 87 & & 26.5& --& 15.7 & & 87& --& 49 & & 15.7& --& 8.8 \\
J2329$+$4743 & & 55& 44& 48 & & 7.5& 6.0& 6.7 & & 10& 6& 33 & & 1.4& 0.8& 4.5 \\
\bottomrule
\end{tabular}
\end{table*}

Full widths at $20\%$ ($W_{20}$) and $50\%$ ($W_{50}$) of the profile peak were calculated. 
In order to improve the precision of the calculated pulse limits, the number of phase bins in the profiles was increased by a factor of 10,000 with a linear interpolation. 
Full widths were obtained as phase differences between the first and last intersects of the profile with a line at $20$ and $50\%$ of the peak, respectively.
Only the phase bins containing the pulse were selected in order to avoid spurious noise peaks to bias the result for low S/N profiles.
The pulsar duty cycles $\delta$ were obtained by calculating the ratio between the width W and the pulsar period.
The results are summarized in Table~\ref{tab:profiles}.

The obtained duty cycles are usually below 10\%. 
The only exceptions are PSRs~J1848+1516 and, most notably, J2051+1248, whose peak occupies more than a quarter of the pulse profile at $149$\,MHz.
For most of the pulsars, the duty cycle decreases with increasing observing frequency.
This is a common behavior explained for example by the radius-to-frequency-mapping model \citep{rud75,cor75}.
However, a few pulsars have wider peaks at higher frequencies.
This could be due to the appearance of additional components in the higher-frequency profile, e.g. in PSRs~J1236$-$0159, J1635$+$2332 and J1848$+$1516, similar to the exceptional cases reported by \citet{pil16}.

The pulse profile of PSR J1848$+$1516 is remarkably different at different observing frequencies.
The phase of the main peak at $149$\,MHz corresponds to the phase of a secondary component in the $1532$-MHz profile.
Instead, no features are present in the $149$-MHz profile coincident with the main peak in the $1532$-MHz profile, although it could become the secondary component shifted at an earlier phase.
In addition, the different components overlap in the $1532$-MHz profile, even though this could be due to the higher noise with respect the $149$-MHz profile.
Finally, the $334$-MHz profile presents only one narrow component clearly detected, coincident with the secondary component in the $149$-MHz profile.
However, only $\sim 25$ of the $816$ pulsar rotations summed to obtain the pulse profile at $334$-MHz contained detectable pulses.

None of the pulse profiles are heavily scattered.
The main peaks of some pulsars, such as PSRs J0115$+$6325, J0139$+$3336, J1236$-$0159, J1635$+$2332 and J1735$+$6320, show a tail at $149$\,MHz that might be consistent with a scattered component.
However, we did not attempt to model the eventual scattering because of its weak effect.

\section{Flux densities and spectral indices}\label{sec:fluxes}
Mean flux densities have been calculated by using the following version of the radiometer equation, obtained by expanding Eqs. 7.1 and A1.21 in \citet{lor04}, after normalizing the pulse profiles.
\[
S_\text{mean} = \frac{T_\text{sky} + T_\text{rec}}{G\sqrt{2t\Delta f}} \sqrt{\frac{\max(p)}{n\max(p)-\sum_{1}^{n}p_i}}\sum_{1}^{n}p_i,
\]
where $p$ is the signal amplitude in a phase bin, $n$ is the total number of phase bins in the pulse profile, $T_\text{sky}$ is the sky temperature, $T_\text{rec}$ is the receiver noise temperature, $G$ is the telescope gain, $t$ is the integration time and $\Delta f$ is the effective bandwidth free from RFI.
Pulse width and period have been expressed in units of phase bins so that $P = n$.
For Lovell telescope observations, we assumed a gain $G=1$\,K\,Jy$^{-1}$ and a system temperature $T_{sys}=25$\,K for the $1532$-MHz receiver \citep{bas16} and $T_{sys}=50$\,K for the $334$-MHz receiver.
The typical RFI environment at the Lovell telescope is estimated to leave $\sim 60$ and $75\%$ of clean data at $334$ and $1532$\,MHz, respectively.
For the NRT telescope, we used  a gain $G=1.6$\,K\,Jy$^{-1}$ and a system temperature $T_{sys}=35$\,K \citep{the05}, while the RFI environment is estimated to leave $\sim 50\%$ of clean data.
Since the LOFAR gain depends on source elevation, number of functional antennas and observing frequency, LOFAR observations were calibrated following the procedure described by \citet{nou15} and \citet{kon16}.
The sky temperature $T_{sky}$ was calculated for each source and frequency by using the $408$-MHz sky map by \citet{has82} and a spectral index $\alpha \sim -2.55$ \citep{moz17}.
Observations too contaminated by RFI or with pointing positions too inaccurate that were refined at a later stage were excluded from the analysis.
LOFAR observations were split into two frequency sub-bands to increase the constraint on the spectral index.
This was not possible for PSRs J1933$+$5335 because the source was too weak and the full LOFAR bandwidth was thus used to calculate the flux of this pulsar.
We decided to exclude PSR J0139$+$3336 from the analysis because its nulling fraction is too extreme to obtain reliable flux values in the duration of our observations.
PSR J2051$+$1248 was barely visible at $334$\,MHz and the signal was too weak to calculate a flux density.

\begin{figure*}
\centering
\includegraphics{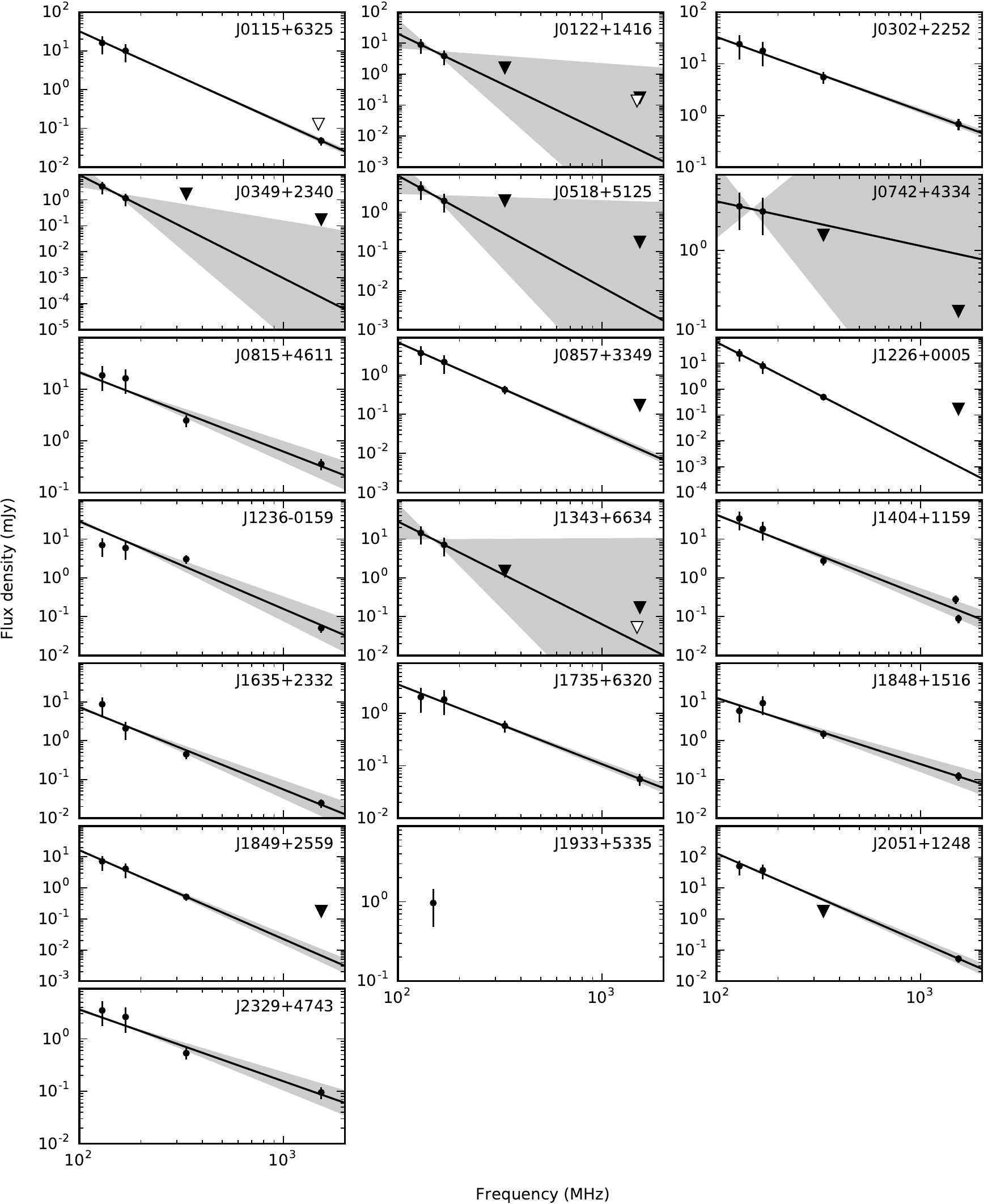}
\caption{
Mean flux densities (dots) and fitted power-law spectra (lines) reported in Table~\ref{tab:flux} for the different pulsars and observing frequencies.
Triangles represent upper limits considering a fiducial value for detection of $\text{S/N}>5$, with filled triangles for Lovell observations and empty triangles for NRT observations.
Shadowed regions represent 1-$\sigma$ uncertainties on the spectral indices referenced to $149$\,MHz.
}
\label{fig:flux}
\end{figure*}

For each pulsar, all Lovell and NRT observations from the same receiver have been added together by using \textsc{psradd} from \textsc{psrchive}, which weights the data by the observation duration and RFI level.
The mean flux density was then calculated from the resulting integrated profile.
This is expected to effectively remove the effect of diffractive scintillation observed at higher frequencies, which caused the flux to vary up to 5 times the average value for different observations of the same pulsar.
It was not possible to follow the same method for LOFAR observations since the number of active antennae varied between observations.
However, since the resulting array sensitivity is not expected to be significantly affected and the RFI level and observation duration are approximately constant, we obtained the mean flux density by averaging the flux density measured in different observations.
The resulting values of flux densities are reported in Table~\ref{tab:flux} and shown in Fig.~\ref{fig:flux}.
We calculated a spectral index or upper limits in case of non-detections.
Due to the small number of measurements and large uncertainties, all the flux values have been fitted with a single power-law.
However, a spectral turn-over is sometimes observed in pulsars at LOFAR frequencies \citep[e.g.][]{bil16} and this could be the case of e.g. PSRs J1236$-$0159 and J1848$+$1516.
The values of the spectral indices obtained vary significantly (Table~\ref{tab:flux}) but the spectra of most pulsars are steeper than the average pulsar population \citep[$\alpha \approx -1.4$,][]{bat13}.
In the cases of a non-detection with one of the telescopes, upper limits were derived by assuming that a $\text{S/N}>5$ is needed to confidently detect a source.

\begin{table*}
\begin{threeparttable}
\centering
\caption{Mean flux densities $S$ measured at different frequencies (indicated in units of MHz as subscripts) and the inferred spectral indices $\alpha$.
Flux density values have been fitted with a single spectral index. The last column reports the offset between the center of the telescope beams and the position refined with timing models.
}
\label{tab:flux}
\begin{tabular}{llllllll}
\toprule
PSR & S$_{129}$ & S$_{168}$ & S$_{334}$ & S$_{1484}$ & S$_{1532}$ &  $\alpha$ & offset \\
 & mJy & mJy & mJy & mJy & mJy & & arcmin \\
\midrule
J0115$+$6325 &     16(8) &         10(5) &        -- &      < 0.1 &    0.05(1) &  -2.38(6) &     2.3 \\
J0122$+$1416 &      9(4) &          4(2) &     < 1.6 &      < 0.1 &      < 0.2 &     -3(3) &     0.6 \\
J0302$+$2252 &    20(10) &         18(9) &      5(1) &         -- &     0.7(2) &  -1.43(6) &     1.6 \\
J0349$+$2340 &      3(2) &        1.1(6) &     < 1.6 &         -- &      < 0.2 &     -4(3) &     0.8 \\
J0518$+$5125 &      4(2) &          2(1) &     < 2.0 &         -- &      < 0.2 &     -3(3) &     1.9 \\
J0742$+$4334 &      4(2) &          3(2) &     < 1.6 &         -- &      < 0.2 &     -1(3) &     0.0 \\
J0815$+$4611 &     18(9) &         16(8) &    2.5(6) &         -- &    0.36(9) &   -1.5(2) &     0.4 \\
J0857$+$3349 &      4(2) &          2(1) &    0.4(1) &         -- &      < 0.2 &  -2.30(8) &     0.9 \\
J1226$+$0005 &    20(10) &          8(4) &    0.5(1) &         -- &      < 0.2 &  -4.05(2) &     2.0 \\
J1236$-$0159 &      7(3) &          6(3) &    3.0(8) &         -- &    0.05(1) &   -2.3(4) &     3.9 \\
J1343$+$6634 &     14(7) &          7(4) &     < 1.5 &      < 0.1 &      < 0.2 &     -3(3) &     1.4 \\
J1404$+$1159 &    30(20) &         18(9) &    2.7(7) &    0.28(7) &    0.09(2) &   -2.1(2) &     1.5 \\
J1635$+$2332 &      9(4) &          2(1) &    0.4(1) &         -- &   0.025(6) &   -2.1(3) &     0.8 \\
J1735$+$6320 &      2(1) &        1.8(9) &    0.6(1) &         -- &    0.06(1) &  -1.52(8) &     0.6 \\
J1848$+$1516 &      6(3) &          9(5) &    1.5(4) &         -- &    0.12(3) &   -1.7(2) &     0.4 \\
J1849$+$2559 &      7(4) &          4(2) &    0.5(1) &         -- &      < 0.2 &   -2.9(2) &     1.1 \\
J1933$+$5335 &       \multicolumn{2}{c}{1.0(5)$^\text{a}$} &   -- &         -- &        -- &        -- &     3.0 \\
J2051$+$1248 &    50(30) &        40(20) &     < 1.8 &         -- &    0.05(1) &   -2.9(2) &     1.1 \\
J2329$+$4743 &      3(2) &          3(1) &    0.5(1) &         -- &    0.10(2) &   -1.4(2) &     0.8 \\

\bottomrule
\end{tabular}
\begin{tablenotes}
\item[a] Value referenced to a central frequency of $149$\,MHz.
\end{tablenotes}
\end{threeparttable}
\end{table*}

For all the telescopes, the random error on flux densities for specific pulsars and frequencies was obtained as the standard error if more than 9 observations were available.
The systematic error due to uncertain estimates of the telescope gain, temperature and average RFI environment is estimated to be $50\%$ of the flux density value for LOFAR measurements \citep{kon16} and $25\%$ for NRT and Lovell.
The resulting uncertainties on the flux density were estimated by using the largest of the two errors.
In addition to these estimated uncertainties, however, there are potentially significant errors on the flux density values that are not accounted for.
Most important, the observations were acquired before timing models were available and thus approximate source positions were used.
The offset of the refined position from the beam center implies actual flux densities somewhat higher than those reported, and consequently steeper spectra.
The uncertainties on the initial positions were of the order of arcminutes, as discussed in \textsection\ref{sec:observations}.
Given the FWHM of LOFAR beams ($\sim 3.5$\,arcmin) the offset from the center of telescope is important for some pulsars.
For comparison, the FWHM of Lovell beams are $\sim 40$ and $9$\,arcmin at $334$ and $1532$\,MHz, respectively.
Correcting for the complex beam shape of LOFAR \citep[e.g.][]{obr15} is difficult and the simple approach used by \citet{san18} to model the beam as a $\sinc^2$ function proved insufficient for this study.
Moreover, ionospheric effects are expected to cause a jitter of LOFAR beams up to $\sim 1$\,arcmin.
Therefore, we do not attempt to correct for the positional offset and caution the reader that the values reported for LOFAR fluxes and spectral indices are indicative and should not be used for detailed studies.
We report the offset between the beam center and the refined position of the pulsars in the last column of Table~\ref{tab:flux}.

NRT observations of PSR J1404$+$1159 were independently calibrated by regularly observing known calibration sources.
This method could not be applied to Lovell data because calibration sources were not observed nor to LOFAR data because of the dependency of the telescope gain on the source position.
The average flux density obtained for PSR J1404$+$1159 with NRT through calibration sources was $0.7(2)$\,mJy, $\sim 1.5\sigma$ away from the value of $0.28(7)$\,mJy obtained through the radiometer equation and reported in Table~\ref{tab:flux}.
We could not identify a clear reason for this discrepancy and it could originate from the systematic errors described above.
In addition, \citet{bri18} report a flux density for the source of $4.3(9)$\,mJy at $327$\,MHz and $0.027(5)$\,mJy at $1400$\,MHz using Arecibo. 
While the first value is compatible with our measurement at $334$\,MHz, the second is lower than our estimates at both $1484$ and $1532$\,MHz.
However, the flux density of the pulsar is highly variable at 1.4\,GHz and we measure values between $\sim 0.04$ and $1.1$\,mJy in individual NRT observations calibrated with known sources.

We checked the TIFR GMRT Sky Survey \citep[TGSS;][]{int17} source catalog around the position of the brightest pulsars at $149$\,MHz in our sample but we did not find any counterpart.

\section{Individual sources}\label{sec:variations}

\begin{figure*}
\centering
\includegraphics{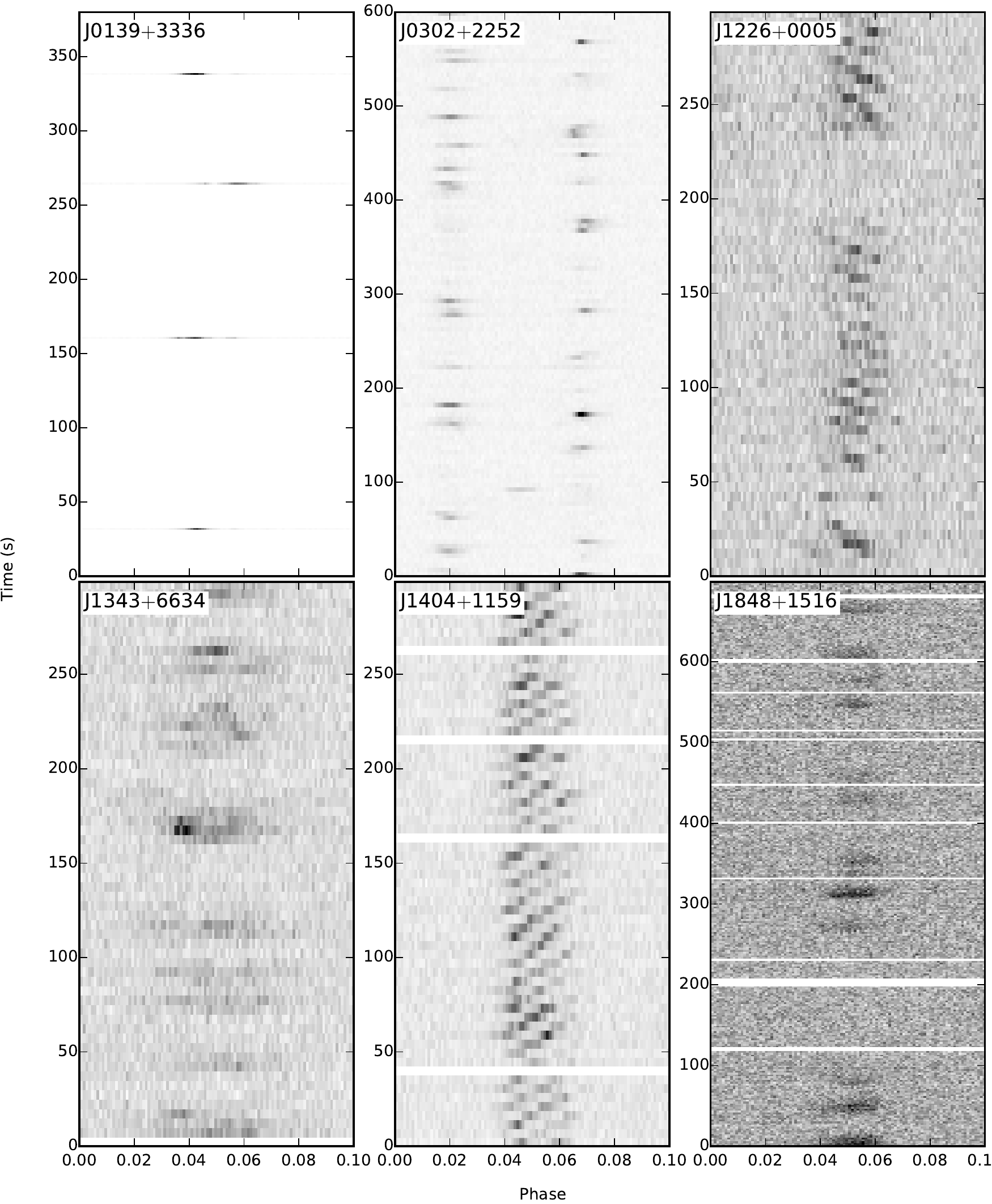}
\caption{Phase-resolved flux density variations over time for six pulsars observed with LOFAR at $149$\,MHz.
The gray-scale is normalized independently for each plot.  
Pulsar names are indicated on the individual panels.
Each time bin contains a single pulsar rotation for PSRs~J0139$+$3336 and J1848$+$1516; time bins are 5\,s for the rest of the sources.
There are $1024$ phase bins over the full phase, $10\%$ of the full rotation is shown here.
Horizontal white stripes indicate data that have been excised to remove RFI.
}
\label{fig:ds}
\end{figure*}

Here we discuss six pulsars of the sample that show sporadic emission or interesting single-pulse behavior.
The flux densities of these sources as a function of rotational phase and time is shown in Fig.~\ref{fig:ds} for LOFAR observations.
Two pulsars show drifting sub-pulses and five are nullers, with PSR~J0139$+$3336 tentatively classified as a RRAT (extreme nuller).
We calculated the nulling fractions of these pulsars following the procedure of \citet{wan07}.
All the observations, with the exception of those relative to PSRs J0139$+$3336 and J1848$+$1516, are averaged every 5 seconds and single pulses are not stored.

After excluding PSR~J0139$+$3336, four pulsars in our sample show nulling fractions $>15\%$.
Therefore, the percentage of nulling pulsars in our sample is more than double the percentage in the total pulsar population, where nulling pulsars are $\lesssim 10$\% \citep{yan14}.
Since the characteristic age of the pulsars in our sample is on average larger than the rest of the population, this could support the evidence found by \citet{rit76} and \citet{wan07} that the nulling fraction is related to the pulsar characteristic age, or it could be due to the long dwell time of LOTAAS observations (one hour each).
The nulling fractions found in our sample (between 15 and 50\%) is large with respect to the rest of the nulling pulsars but not unheard-of \citep[e.g.][]{big92,wan07}.

\subsubsection*{PSR~J0139$+$3336}

\begin{figure}
\centering
\includegraphics{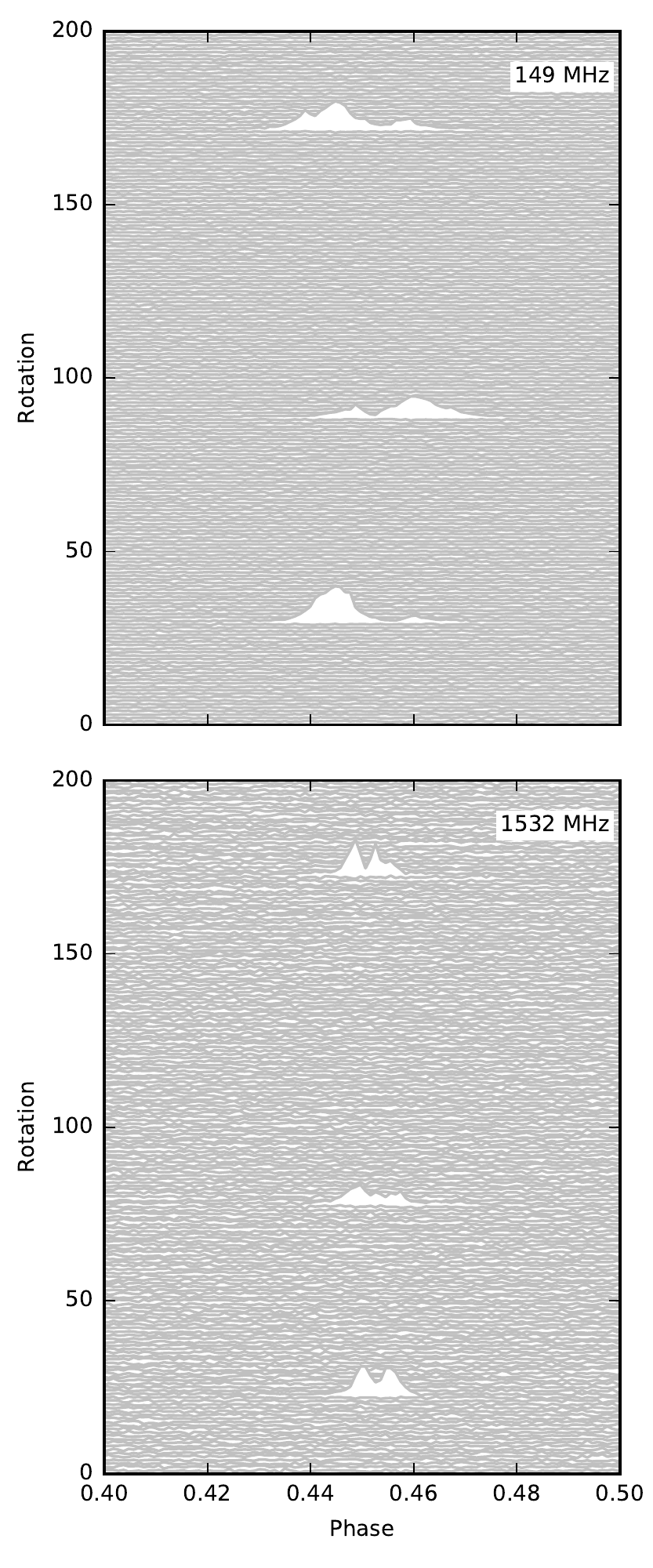}
\caption{Flux density (in arbitrary units) as a function of phase in single rotations of PSR~J0139$+$3336.
The two observations (LOFAR at $149$\,MHz, top, and Lovell at $1532$\,MHz, bottom) are not simultaneous.
The same number of rotations are shown for the two observations.
}
\label{fig:sp}
\end{figure}

The source shows the behavior of a RRAT, with only sporadic single pulses detected.
We stored single-pulse resolved data at both $149$ and $1532$\,MHz.
Pulses are visible in single pulsar rotations separated by minutes.
An example of a few bright pulses is reported in Fig.~\ref{fig:sp}.

Using LOFAR observations, we selected pulses having a $\text{S/N}>10$ at the phase of the main peak in the integrated pulse profile.
This threshold was chosen to select pulses clearly separated from the noise distribution.
A total of 30 pulses were detected in LOFAR observations above this threshold.
An average of $\sim 3$ pulses per observation was detected, implying a rate of one pulse every $\sim 5$ minutes.
Given the small number of detected events, the rate of pulses among different observations was roughly compatible with a Poisson distribution.

The same analysis was repeated for Lovell observations at $1532$\,MHz.
The S/N of the brightest pulses is similar in the two cases.
Also the rate is similar, with a pulse detected every $\sim 5$ minutes with a $\text{S/N}>10$ and a distribution roughly consistent with a Poisson distribution.
The lack of a robust estimate of the source spectral index prevents a more detailed comparison of the pulses at the two frequencies.

A peak in the integrated pulse profile was detected in all of LOFAR and most of Lovell observations containing pulsar rotations where emission could be visually identified. 
After excluding these single rotations, the integrated pulse profile is indistinguishable from noise.

\subsubsection*{PSR~J0302$+$2252}
The flux of this nulling pulsar is highly variable on short timescales for both the peaks in the profile (Fig.~\ref{fig:ds}).
Unfortunately, we did not store single pulses for this source; rather, the flux is averaged every five seconds (about four rotational periods).
Therefore, it is impossible to assess the flux variability over single rotations.
The degree and timescale of variation are similar for the two peaks, with a nulling fraction $\sim 15$\%.
However, the flux density of the two peaks in single sub-integrations is not obviously correlated.

\subsubsection*{PSR~J1226$+$0005}
A null emission lasting $\sim 40$ seconds can be seen for this pulsar in Fig.~\ref{fig:ds} around $200$ seconds after the start of the observation. 
Longer nulls are detected as well, with the pulsar being detected for only the first $\sim 2$ minutes of one 10-minute observation.
The average nulling fraction for the pulsar is $\sim 50\%$.

Fig.~\ref{fig:ds} also reveals drifting sub-pulses for PSR J1226$+$0005. 
With no individual pulses being recorded (Fig.~\ref{fig:ds} shows 5 second, or $\sim2.2$ pulse period averages) it is hard to quantify this further. 
Nevertheless, the drift-rate appears to be variable with the drift-rate being lower as seen at the top of Fig.~\ref{fig:ds} (i.e. the drift bands are steeper) compared to what it is at $\sim 120$ seconds into the observation. 
In addition, the emission appears to wander slightly in pulse phase (e.g. the emission is slightly late $\sim120$ seconds into the observation shown in Fig.~\ref{fig:ds}). 
Individual pulse observations might reveal if the observed variability is related to discrete mode changes, or if the effect is smoother.

\subsubsection*{PSR~J1343$+$6634}

This pulsar shows a nulling fraction $\sim 35$\%.
The source switches between detectable and non-detectable states on a timescale of a few tens of seconds.
This behavior is consistent throughout the different observations.

\subsubsection*{PSR~J1404$+$1159}

During the preparation of this manuscript, the source has also been studied by \citet{bri18} at $327$ and $1400$\,MHz using the Arecibo telescope.
The parameters that they present are in agreement with our measurements. 
We also detect the bright drifting sub-pulses forming the main peak.
The sub pulses are clearly visible in the 5-second long sub-integrations visible in Fig.~\ref{fig:ds}.
The relatively high S/N of the pulsar and detection of the drifting sub-pulses over multiple frequencies could allow detailed studies of the drifting evolution with frequency \citep[e.g.][]{has13}.

\subsubsection*{PSR~J1848$+$1516}

The source switches between detectable and non-detectable states every few tens of rotations, with an average nulling fraction of $\sim 50$\%. 
While the pulsar is relatively active in some observations (as shown in Fig.~\ref{fig:ds}) it is undetected in several 15-minute observations.
Sporadically, a second peak appears trailing the main one, becoming the brightest in three observations.
Only on a very few occasions, a third peak leading the main one has been detected for a few rotations.

\section{Conclusions}\label{sec:conclusions}
We have presented the properties of 20 radio pulsars discovered by the LOFAR telescope as part of the LOTAAS survey.
Since their discovery, the sources have been regularly observed at multiple frequencies using LOFAR, Lovell and NRT telescopes.
This allowed us to calculate the astrometric and rotational parameters of the pulsars.
They have, on average, longer periods and lower spin-down rates than the majority of the pulsar population.
This places the pulsars closer to the death line than the average of the global pulsar population.
It is unclear whether this is a real effect or a selection bias and this will be explored in a subsequent paper using a larger LOTAAS sample.
Integrated pulse profiles were calculated at different frequencies using the obtained timing models.
They are mostly single-peaked and show frequency evolution with a complex behavior in some cases.
Values of mean flux densities at the different observing frequencies have been calculated.
Even keeping in mind the systematic errors present, the resulting spectra are steeper than average for most pulsars.
%However, two pulsars have a spectrum flatter than the average population, with an index $\sim -1.2$.
Five out of the 20 pulsars in the sample are undetectable for more than $15\%$ of time, with PSR J0139$+$3336 tentatively classified as a RRAT.
Two of the pulsars show drifting sub-pulses.

%%%%%%%%%%%%%%%%%%%%%%%%%%%%%%%%%%%%%%%%%%%%%%%%%%

\section*{Acknowledgements}
We thank Anne M. Archibald for her help with the analysis.
DM and JWTH acknowledge funding from the European Research Council under the European Union's Seventh Framework Programme (FP/2007-2013) / ERC Starting Grant agreement nr. 337062 (`DRAGNET').  
JWTH also acknowledges funding from an NWO Vidi fellowship.
DM is a Banting fellow.
JvL acknowledges funding from  the European Research Council under the European Union's Seventh Framework Programme (FP/2007-2013) / ERC Grant Agreement n. 617199 
(`ALERT'), and from Vici research programme `ARGO' with project number 
639.043.815, financed by the Netherlands Organisation for Scientific 
Research (NWO).
This paper is based in part on data obtained with the International LOFAR Telescope (ILT). 
LOFAR \citep{haa13} is the Low Frequency Array designed and constructed by ASTRON. 
It has facilities in several countries. 
These are owned by various parties (each with their own funding sources) and are collectively operated by the ILT
foundation under a joint scientific policy.
The data for this project were taken as part of the LOFAR long-term proposals ``Pulsar Timing with LOFAR" (PI: Verbiest), under proposal codes LC0\_011, DDT0003, LC1\_027, LC2\_010, LT3\_001, LC4\_004, LT5\_003, LPR6\_002, LPR7\_001, LPR8\_001, LC9\_041 and LT10\_004, and ``Additional Timing Observations on LOTAAS Pulsar Discoveries'' (PI: Tan), under proposal codes LC8\_035, LC9\_021 and LT10\_015.
Pulsar research at Jodrell Bank and access to the Lovell Telescope is supported by a Consolidated Grant from the UK's Science and Technology Facilities Council.
The Nan\c{c}ay Radio Observatory is operated by the Paris Observatory,
associated with the French Centre National de la Recherche Scientifique 
(CNRS) and Universit\'{e} d'Orl\'{e}ans.

\bibliographystyle{mnras}
\bibliography{mnras}

% Don't change these lines
\bsp	% typesetting comment
\label{lastpage}
\end{document}